\documentclass[epj]{webofc}
\usepackage[varg]{txfonts}   % Web of Conferences font
\usepackage{xspace}

\usepackage{lineno}
\usepackage{color}
\usepackage{ulem}

\modulolinenumbers[1]

\woctitle{International Conference on New Frontiers in Physics 2013}

\begin{document}

\newcommand{\ppb}{\rm{p-Pb}\xspace}
\newcommand{\pp}{\ensuremath{\rm pp}\xspace}
\newcommand{\ks}{\ensuremath{\rm K_{S}^{0}}\xspace}
\newcommand{\lmd}{\ensuremath{\Lambda}\xspace}
\newcommand{\cme}{\ensuremath{\sqrt{s}}\xspace}
\newcommand{\bg}{\ensuremath{\beta\gamma}\xspace}
\newcommand{\pt}{\ensuremath{p_{\rm{T}}}\xspace}
\newcommand{\p}{\ensuremath{p}\xspace}
\newcommand{\gev}{\ensuremath{{\rm GeV}/c}\xspace}
\newcommand{\pbpb}{Pb--Pb\xspace}
\newcommand{\auau}{Au--Au\xspace}

\newcommand{\twotwo}{\ensuremath{2\rightarrow 2}\xspace}
\newcommand{\raa}{\ensuremath{R_{\rm AA}}\xspace}
\newcommand{\dedx}{\ensuremath{{\rm d}E/{\rm d}x}\xspace}
\newcommand{\mdedx}{\ensuremath{\langle {\rm d}E/{\rm d}x \rangle}\xspace}

\newcommand{\pikp}{\ensuremath{\pi^{\pm}, \rm K^{\pm}, \rm p(\bar{p})}\xspace}
\newcommand{\ptopi}{\ensuremath{({\rm p+ \bar{p}}) / (\pi^{+}+\pi^{-})}\xspace}
\newcommand{\ktopi}{\ensuremath{({\rm K}^{+}+{\rm K}^{-}) / (\pi^{+}+\pi^{-})}\xspace}
\newcommand{\chpi}{\ensuremath{\pi^{+}+\pi^{-}}\xspace}
\newcommand{\chk}{\ensuremath{{\rm K}^{+}+{\rm K}^{-}}\xspace}
\newcommand{\chp}{\ensuremath{{\rm p}+{\rm \bar{p}}}\xspace}

\title{Results on identified particle production in pp, p-Pb and Pb-Pb collisions measured with ALICE at the LHC}

\author{Antonio Ortiz Velasquez (for the ALICE Collaboration)\inst{1}%\fnsep\thanks{\email{antonio.ortiz@nucleares.unam.mx}}
\fnsep\thanks{Now at: Instituto de Ciencias Nucleares, Universidad Nacional Aut\'onoma de M\'exico,  Apartado Postal 70-543, M\'exico Distrito Federal 04510, M\'exico, \email{antonio.ortiz@nucleares.unam.mx}}}

\institute{Lund University, Department of Physics, Division of Particle Physics Box 118, SE-221 00, Lund, Sweden.}

\abstract{%

Using the unique capabilities of the ALICE detectors for particle identification, different measurements have been performed to study the properties of the hot and dense matter created in the \pbpb collisions at $\sqrt{s_{\rm NN}}=$ 2.76 TeV. The analysis of the \ppb data at $\sqrt{s_{\rm NN}}=$ 5.02 TeV reveals that the suppression of high \pt hadrons observed in heavy nuclei collisions can not be explained as due to initial state effects. The systems created in the \ppb collisions do not show evidence of jet quenching but, surprisingly, exhibit characteristics of flow. In this paper a review of the main results on identified particle production measured in different colliding systems is presented, data are also compared to models.

}
\maketitle
\section{Introduction}
\label{intro}

In ultra relativistic heavy ion collisions a new form of matter characterized by the deconfined state of quarks and gluons is formed~\cite{Bass:1998vz}. The hot and dense matter rapidly expands and reduces its temperature. The properties of the medium can be extracted from the study of the final state observables like abundance of particles (pions, kaons,...), transverse momentum (\pt) spectra, correlations, etc~\cite{Adcox:2004mh}. Due to the complexity of the final states, one needs to perform analogous measurements in smaller systems like those created in \pp or \ppb collisions, in order to disentangle the (genuine) initial and final states effects.

In vacuum, the \pt distributions of hadrons contain information sensitive to different physics. Most of the particles are produced  at low \pt ($<2$ GeV/$c$), where an effective theory for describing the non perturbative QCD regime is missing, instead phenomenological models are commonly used. Hence, experimental data are valuable inputs for improving the models. At larger transverse momentum particles belong mainly to jets which originate from hard scatterings and, due to the energy scale of the system, perturbative QCD can be applied. In heavy ion collisions the low to intermediate transverse momentum ($\pt<8$ GeV/$c$) spectra of identified hadrons are excellent tools to study the properties of the medium, namely, hydrodynamical flow and possible new hadronization mechanisms like quark-recombination~\cite{Fries:2008hs}. At larger \pt, the modification of the fragmentation due to the medium can be explored~\cite{Sapeta:2007ad,rene:1}. 

Nucleon-nucleus collisions have been used as control experiments to study the  cold nuclear matter effects. However, recently the experiments at LHC and RHIC have showed that these systems exhibit interesting features which are reminiscent to collective effects observed in heavy nuclei collisions~\cite{ABELEV:2013wsa,tagkey201425}. In the following sections, some results on identified particle production measured in different colliding systems are presented.

\section{ALICE detector}
\label{sec-1}

\begin{figure}[htbp]
  \centering
 
  \includegraphics[keepaspectratio, width=0.49\columnwidth]{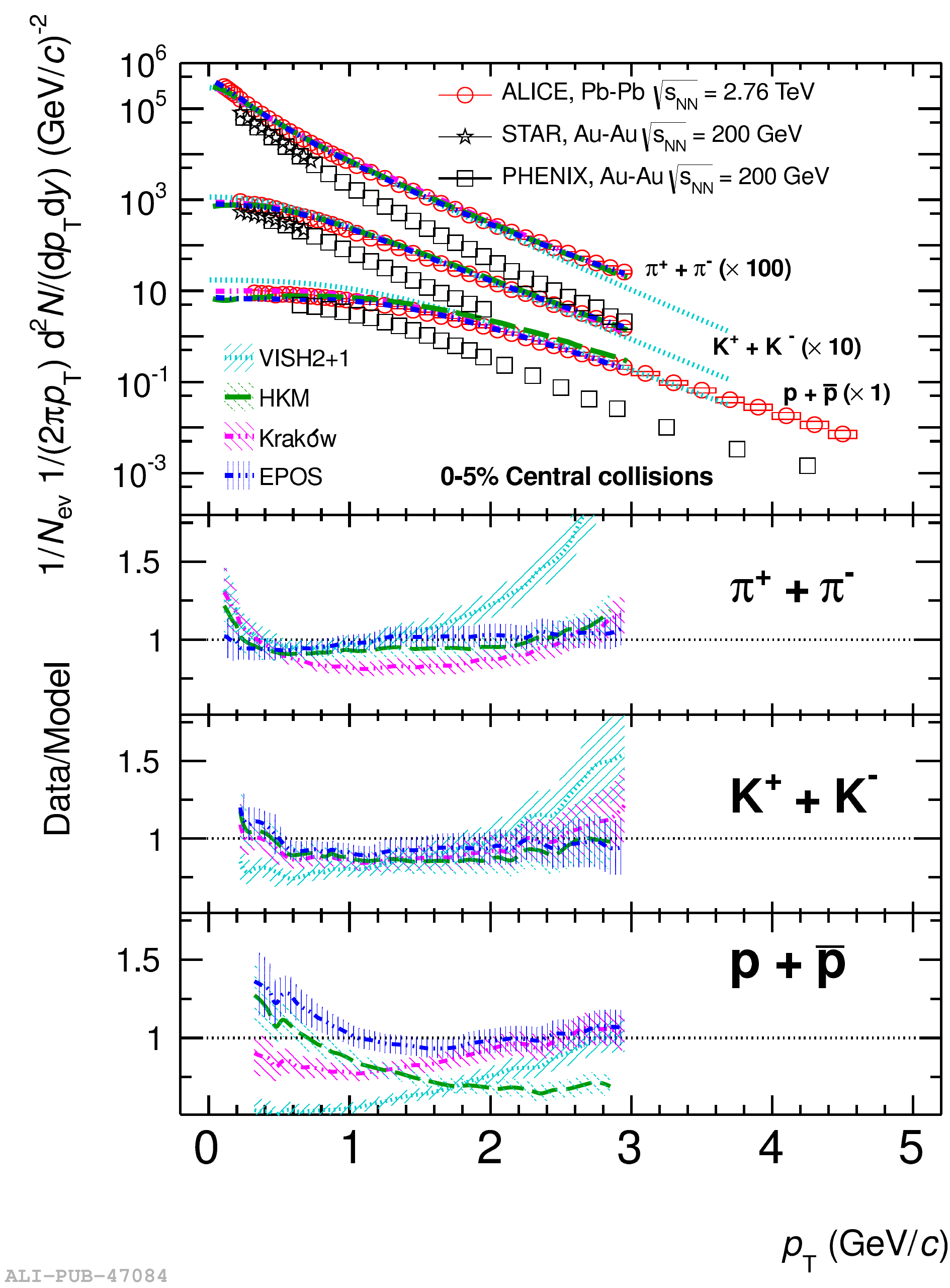}
  \includegraphics[keepaspectratio, width=0.49\columnwidth]{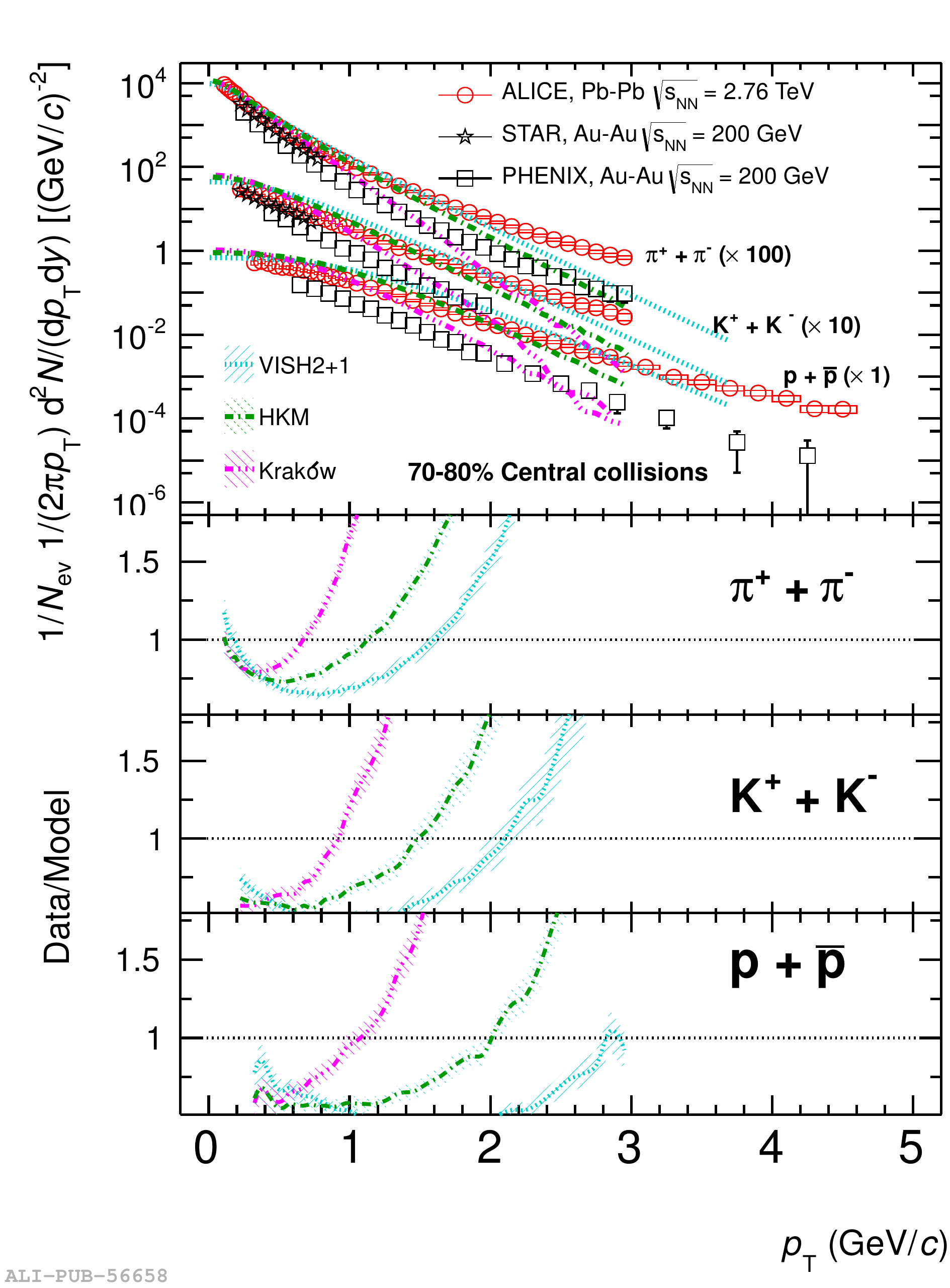}
  \caption{Transverse momentum spectra of charged pions, kaons and (anti)protons measured in central (left) and peripheral (right) \pbpb collisions at $\sqrt{s_{\rm NN}}=$ 2.76 TeV. The results are compared with measurements at RHIC and with hydrodynamical models.}  
  \label{fig:1}
\end{figure}

The results presented here are based on the data collected by ALICE. The Inner Tracking System (ITS) and the Time Projection Chamber (TPC) are used for vertex finding and tracking. The minimum bias trigger was based on the signals from forward scintillators (VZERO) and, in \pp collisions, the silicon pixel detector. For \pbpb collisions the analysis is performed in centrality classes which are determined from the measured amplitude in the VZERO detectors and using simulations based on a Glauber model~\cite{Abelev:2013qoq}. In \ppb collisions the centrality determination is challenging because the correlation between collision geometry and the charged particle multiplicity is very weak, therefore the results are reported in multiplicity event classes based on the amplitude of the signal of VZEROA detector ($2.8<\eta<5.1$)~\cite{tagkey201425} which corresponds to the $\rm Pb$ side.

Particle identification (PID) in ALICE is done at mid-rapidity over a wide range of transverse momentum ($p_{\rm T}$) using different detectors. For $p_{\rm T}$ between $\sim 100$ MeV/$c$ up to 3-4 GeV/$c$ (anti)protons, charged pions and kaons can be separated through the measurement of the specific energy loss (${\rm d}E/{\rm d}x$) in gas (silicon) with TPC (ITS) and time of flight (TOF). The identification  can be extended to higher $p_{\rm T}$ by using a Cherenkov detector (HMPID) in a limited rapidity range. For $3 < p_{\rm T} < 20$ GeV/$c$, statistical PID is possible thanks to the relativistic rise of the ${\rm d}E/{\rm d}x$ in the TPC~\cite{OrtizVelasquez:2012te,Abelev:2014laa}. $\Lambda$ and $\rm K_{S}^{0}$ can be identified using their characteristic weak decay topologies~\cite{PhysRevLett.111.222301}.

\section{Results}
\label{sec-2}

The transverse momentum distributions of charged pions, kaons and (anti)protons measured by ALICE in \pbpb collisions are shown in Fig.~\ref{fig:1}. Results are presented for summed charged particles since the spectra for positive and negative particles are compatible within systematic uncertainties. The spectral shapes exhibit a change with the collision centrality, namely, in the most central collisions (0-5\%) the spectra are harder than those measured in peripheral events (70-80\%). The effect increases with increasing hadron mass, which can be attributed to the radial flow. In addition, from the blast-wave~\cite{PhysRevC.48.2462} analysis the collective flow measured at the LHC is found to be $\approx$10\% higher than at RHIC~\cite{Abelev:2013vea}. The spectra are also compared with hydrodynamical calculations in which the inclusion of the hadronic phase improves agreement with data. For this purpose HKM~\cite{Karpenko:2012yf} and EPOS~\cite{Werner:2012xh} use the hadronic cascade model UrQMD. The viscous hydrodynamic model VISH2+1~\cite{Song:2010aq} does not contain a similar process/implementation. In general the most central collisions are well described by models but they deviate from data going to the more peripheral events, this feature indicates the limit of the hydrodynamical models. The mid-rapidity \pt integrated particle yields can be extracted using individual fits of the blast wave functions to the \pt spectra. Then, the \pt integrated particle ratios can be compared to predictions from thermal models. For central \pbpb collisions the \pt integrated \ptopi has been found to be in disagreement with predictions assuming a baryochemical potential $\mu_{B}=1$ and chemical freeze-out temperature, $T_{\rm ch}=$ 164 or 170 MeV~\cite{Abelev:2012wca}.  This could be attributed to the annihilation process (the issue is still open).

\begin{figure}[htbp]
  \centering
 
  \includegraphics[keepaspectratio, width=0.445\columnwidth]{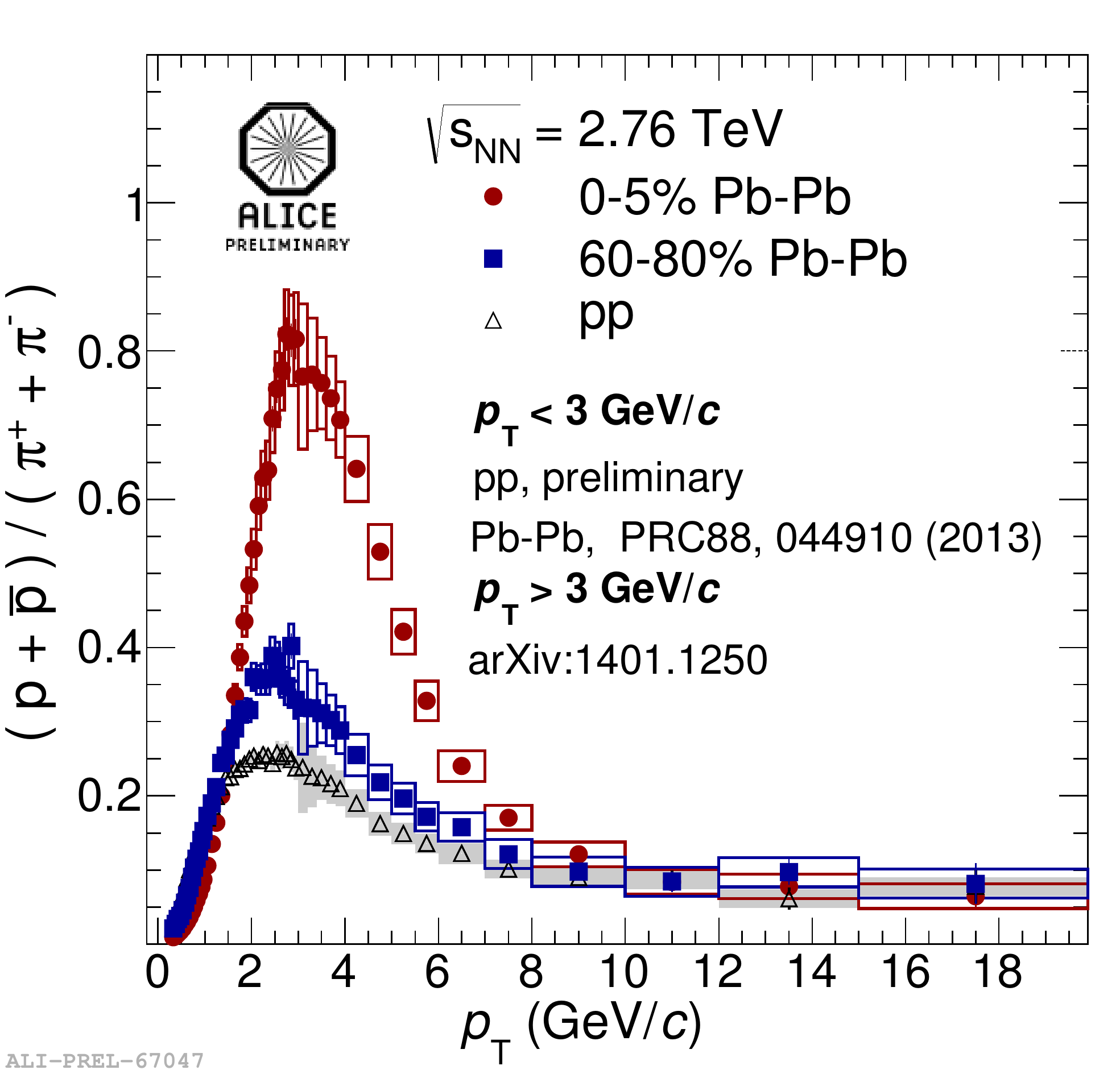}
  \includegraphics[keepaspectratio, width=0.545\columnwidth]{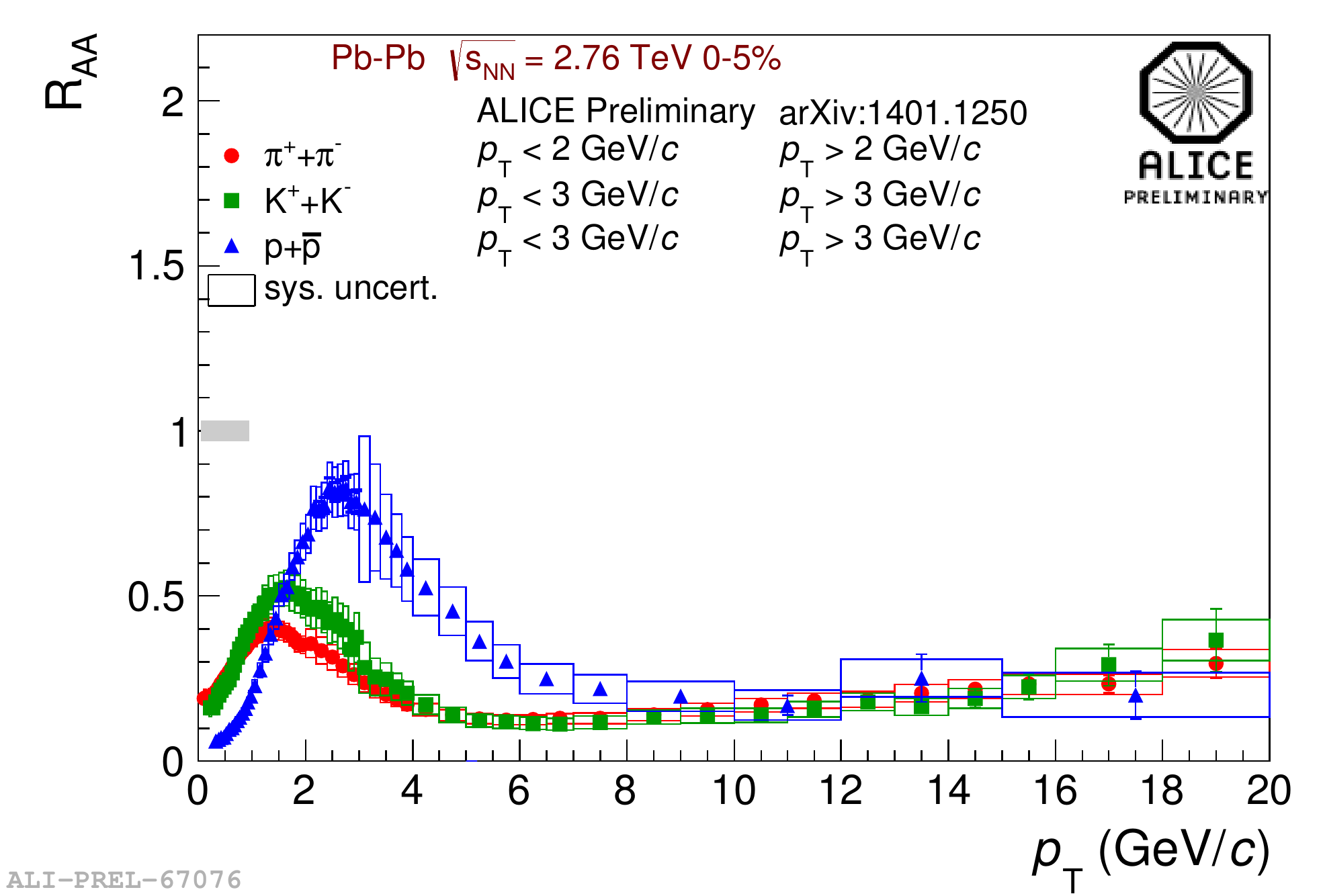}
  \caption{Proton-to-pion ratio as a function of \pt measured in \pp and \pbpb collisions (left). Nuclear modification factor, \raa, as a function of \pt for charged pions, kaons and (anti)protons, the results are shown for central \pbpb collisions (right). The statistical and systematic uncertainties are plotted as vertical error bars and boxes, respectively. }  
  \label{fig:2}
\end{figure}

The proton-to-pion ratio as a function of \pt is plotted in Fig.~\ref{fig:2}, the results are presented for \pbpb collisions in different centrality classes and compared to \pp. Going from peripheral to the more central events the particle ratio exhibits a depletion at low \pt ($<2$ GeV/$c$) and an enhancement at intermediate \pt (2-8 GeV/$c$). The size of the enhancement is $\approx$ 0.8 for 0-5\% collision centrality, this value is compatible to those reported by PHENIX and STAR in central \auau collisions at $\sqrt{s_{\rm NN}}=$ 200 GeV~\cite{raapid:rhic:3,raapid:rhic:2}. The effect can be due to two mechanisms, the radial flow and/or quark recombination. However, so far, there is not a satisfactory full explanation.
The modification of the fragmentation due to the medium can be studied by looking at the particle composition at higher \pt ($>$ 8 GeV/$c$), the baryon-to-meson ratios measured for all the centrality classes and for \pp collisions are consistent within the systematic uncertainties. This indicates that the medium effects, if there, are not changing the particle composition any more at high \pt. The same conclusion is obtained from the  $\Lambda/{\rm K}_{S}^{0}$ ratio~\cite{PhysRevLett.111.222301}.

To study the jet quenching effects, {\it i.e.,} the partonic energy loss due to the interaction of the probe with the medium, the nuclear modification factor, \raa, is measured. This is defined as follows:
\begin{equation}
R_{\rm AA} =  \frac{{\rm d}^{2}N_{\rm ch}^{\rm AA}/{\rm d}\eta{\rm d}\pt}{\langle N_{\rm coll} \rangle {\rm d}^{2}N_{\rm ch}^{\rm pp}/{\rm d}\eta{\rm d}\pt},
\end{equation}
where the \pt spectrum measured in nuclei collisions is normalized to that measured in \pp collisions via the $\langle N_{\rm coll} \rangle$ (number of binary collisions) scaling. What is observed in Fig.~\ref{fig:2} is that at high \pt the charged pions, kaons and (anti)protons are equally suppressed within systematic uncertainties. This suggests that the chemical composition of leading particles from jets in the medium is similar to that of jets produced in vacuum~\cite{Abelev:2014laa}.

\begin{figure}[htbp]
  \centering
 
  \includegraphics[keepaspectratio, width=0.32\columnwidth]{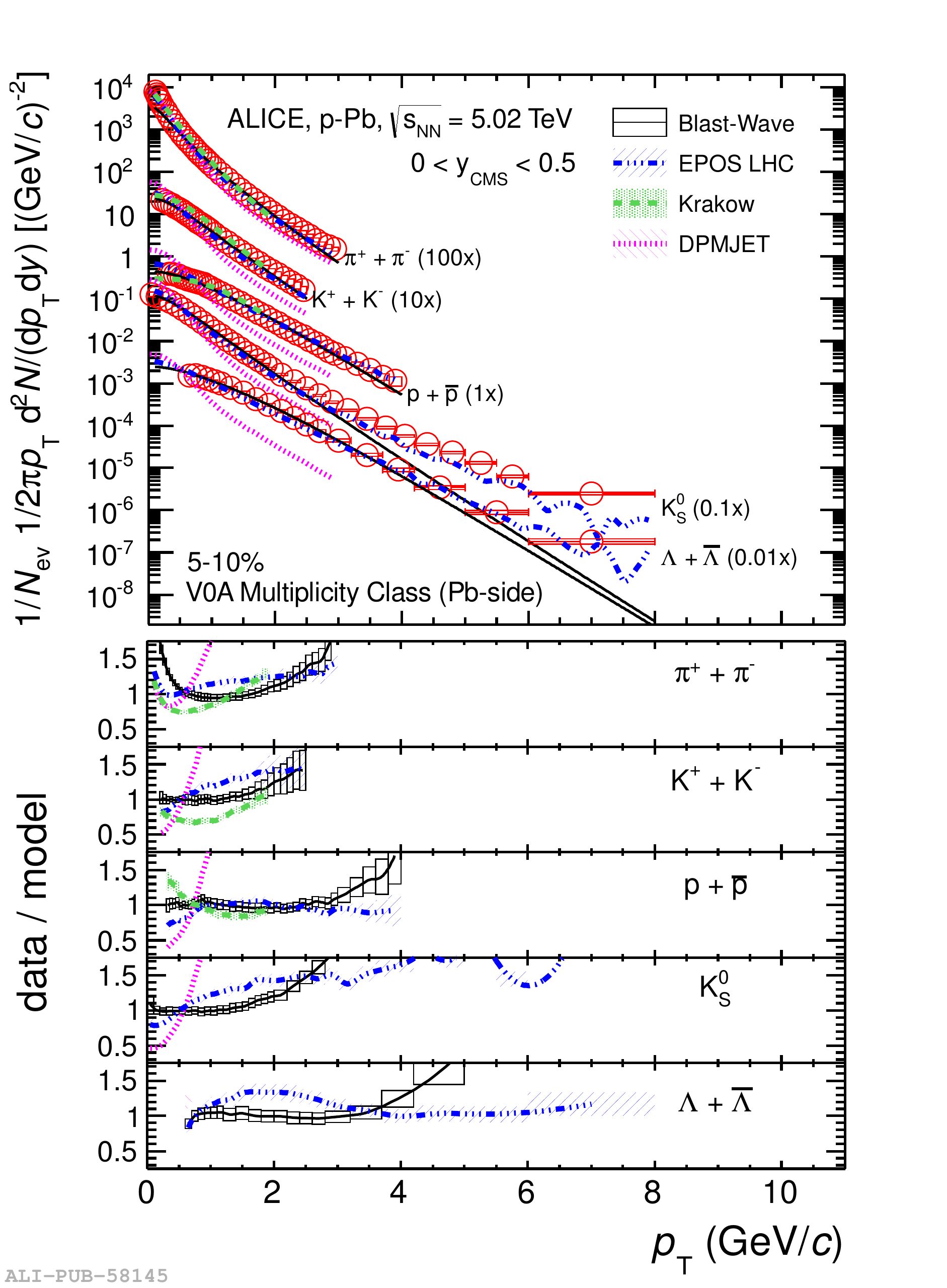}
  \includegraphics[keepaspectratio, width=0.66\columnwidth]{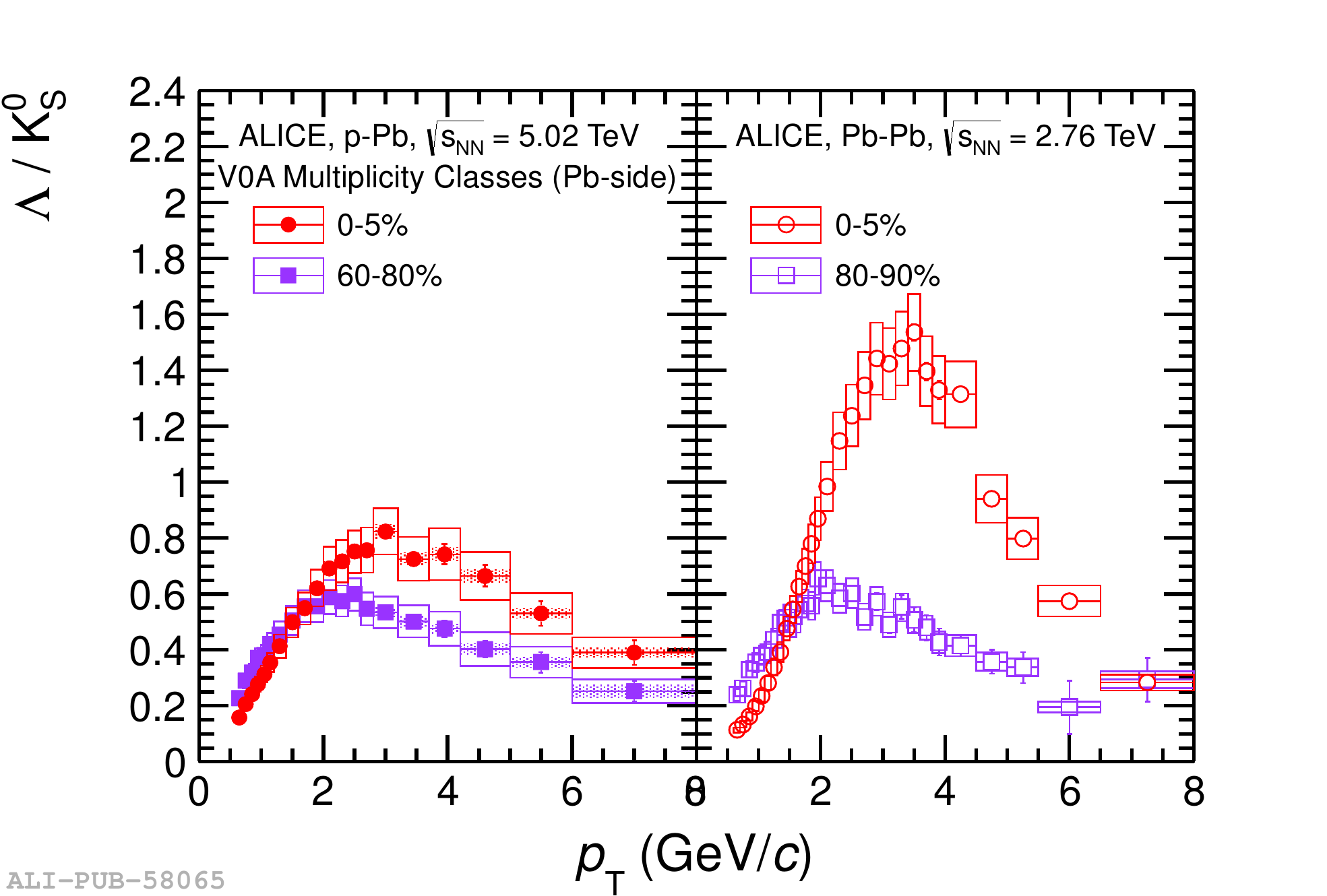}
  \caption{ Transverse momentum spectra of different particle species measured in high multiplicity \ppb collisions, data are compared with models (left). Baryon to meson ratios measured in \ppb and \pbpb collisions, $\Lambda/{\rm K}_{S}^{0}$ vs. \pt are presented for two extreme multiplicity (centrality) classes (right). }  
  \label{fig:3}
\end{figure}

The suppression of the high \pt hadrons in \pbpb collisions is due to final state effects, this conclusion is supported from the measurement of the nuclear modification factor in \ppb collisions which gives a $R_{\ppb}$ consistent with unity at high \pt~\cite{PhysRevLett.110.082302}. However, features like the double ridge structure have been observed in high multiplicity \ppb events suggesting the presence of collective effects~\cite{Abelev:2012ola}. The transverse momentum spectra of identified hadrons exhibit a pattern which is reminiscent to the effects due to the collective expansion of a medium. For example, the spectra become harder with increasing multiplicity, the hardening is more pronounced for protons than for pions. For high multiplicity events, Fig.~\ref{fig:3} (left) shows the comparison of the \pt spectra with models. The general feature is that models including hydro do the best job of describing the data, in contrast, DPMJET deviates quite a lot from data. The latter is a QCD inspired model which describes well the ${\rm d}N/{\rm d}\eta$ distribution~\cite{PhysRevLett.110.032301}. The baryon to meson ratios for two extreme multiplicity (centrality) classes are plotted in Fig.~\ref{fig:3} (right) for \ppb (\pbpb) collisions. Both systems give a $\Lambda/{\rm K}_{S}^{0}$ ratio which shows a depletion (an enhancement) at low (intermediate) \pt when going from low to high centrality or multiplicity.

\begin{figure}[htbp]
  \centering
 
  \includegraphics[keepaspectratio, width=0.75\columnwidth]{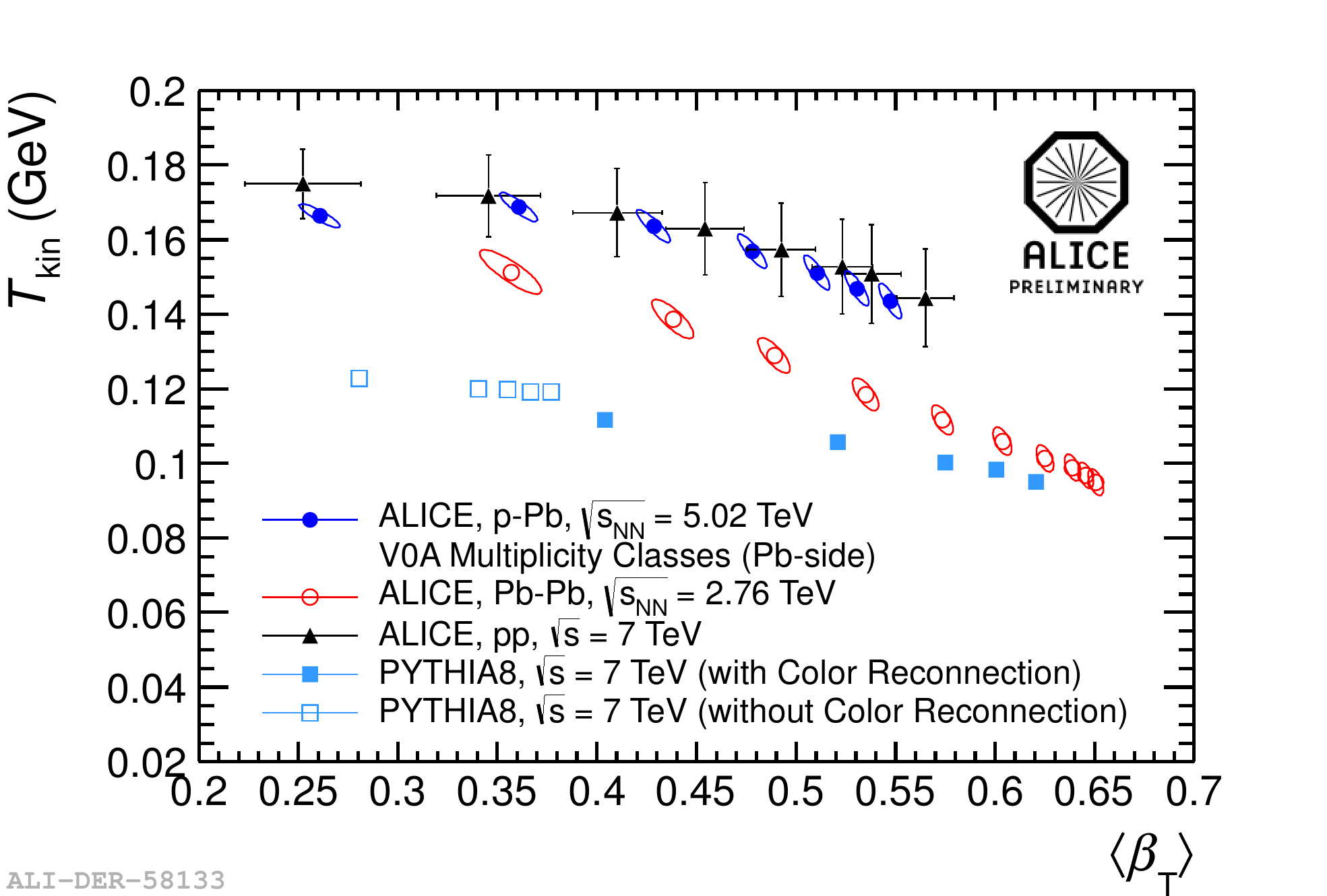}
  \caption{ Comparison of the results from the blast-wave analysis applied to all available systems: \pp, \ppb and \pbpb collisions. The spectral shape analysis was also implemented in MC simulations, the results obtained using Pythia 8 tune 4C are shown.  Charged-particle multiplicity increases from left to right. }  
  \label{fig:4}
\end{figure}

To characterize the evolution of the spectral shapes with the multiplicity the blast-wave analysis is implemented. This tool allows to compare the results obtained in different colliding systems using a small set of parameters: the kinetic freeze out temperature, $T_{\rm kin}$, and the average transverse velocity, $\langle \beta_{\rm T} \rangle$. Figure~\ref{fig:4} shows $T_{\rm kin}$ vs. $\langle \beta_{\rm T} \rangle$ measured in \pp, \ppb and \pbpb collisions. All the systems present a qualitatively similar behaviour; $T_{\rm kin}$ ($\langle \beta_{\rm T} \rangle$) decreases (increases) with increasing multiplicity suggesting presence of flow in all systems. Surprisingly, Pythia 8 tune 4C~\cite{pythia8:1}, which does not include any  hydro collectivity, produces a qualitatively similar pattern. This is a consequence of a flow-like effect due to the ``communication'' among the outgoing partons produced in different (semi) hard scatterings within the same hadron-hadron interaction via color reconnection~\cite{Ortiz:2013yxa}.

\section{Conclusions}
\label{sec-4}

A summary of the main results on the identified particle production measured by ALICE has been presented. The transverse momentum spectra are qualitatively well reproduced by hydrodynamical models at low \pt in central \pbpb collisions, while, for more peripheral events the agreement between data and models is worse. Charged pions, kaons and (anti)protons at high \pt are equally suppressed, the suppression is the largest for central \pbpb collisions. The nuclear modification factor measured in \ppb collisions for inclusive charged particles is consistent with unity for \pt larger than 2 GeV/$c$. This observation supports the idea that the suppression observed in \pbpb data is a final state effect. The \ppb data do not exhibit suppression of high \pt hadrons but give a strong hint of collectivity. Even high multiplicity \pp collisions exhibit flow signatures. The origin of this effect is not well established, but qualitatively similar effects are obtained both in hydrodynamical models and Pythia 8. High multiplicity \pp events generated with Pythia 8 exhibit features that are reminiscent of collective effects although no hydro is present in the generator. The observed effect is attributed to multi-parton interactions and subsequent color reconnection. This opens interesting prospects to investigate the possible existence of similar contributions in heavy ion collisions.

\bibliography{apssamp}{}
\bibliographystyle{unsrt}

\end{document}